\documentclass[a4paper,12pt]{article}
\usepackage{graphicx}
\usepackage{geometry}
\usepackage{booktabs}
\usepackage{multirow}
\usepackage{adjustbox}
\usepackage{xcolor}
\usepackage{array}
\usepackage{float}
\usepackage{authblk}
\usepackage[utf8]{inputenc}
\DeclareUnicodeCharacter{2212}{\textendash}
\title{\textbf{Age Effects on Decision-Making, Drift Diffusion Model}}
\author[1]{Zahra Kavian}
\author[1]{Kimia Hajisadeghi}
\author[2]{Yashar Rezazadeh Shahir}
\author[2]{Mehrbod Faraji}
\author[3]{Reza Ebrahimpour}
\affil[1]{Department of Electrical Engineering, Sharif University of Technology,Tehran, Iran.}
\affil[2]{Department of Computer Engineering, Shahid Rajaee Teacher Training University,Tehran, Iran.}
\affil[3]{Center for Cognitive Science, Institute for Convergence Science and Technology, Sharif University of Technology, Tehran, Iran.}

\date{ }
\begin{document}
\maketitle
\section*{Abstract}
Training can improve human decision-making performance. After several training sessions, a person can quickly and accurately complete a task. However, decision-making is always a trade-off between accuracy and response time. Factors such as age and drug abuse can affect the decision-making process. This study examines how training can improve the performance of different age groups in completing a random dot motion (RDM) task. The participants are divided into two groups: old and young. They undergo a three-phase training and then repeat the same RDM task. The hierarchical drift-diffusion model analyzes the subjects' responses and determines how the model's parameters change after training for both age groups. The results show that after training, the participants were able to accumulate sensory information faster, and the model drift rate increased. However, their decision boundary decreased as they became more confident and had a lower decision-making threshold. Additionally, the old group had a higher boundary and lower drift rate in both pre and post-training, and there was less difference between the two group parameters after training.

\section*{Introduction}
	
	
Mathematical models, such as the diffusion model (DM), are commonly used to understand cognitive processes and analyze observed behavior. DM parameters, including boundary separation, non-decision time, and drift rate, help to explain decision-making behavior. For example, decision caution refers to the amount of information examined before making a decision, measured by the boundary separation parameter. The larger the boundary separation, the more careful one should be. Non-decision time refers to the time involved in encoding stimuli, which includes sampling processes, accumulating evidence, and executing motor responses during decision-making.

Mischa von Krause \cite{von2022mental} used a Bayesian diffusion model in 2022 to analyze raw response time data from a large dataset consisting of 1.2 million participants. The study showed that response time starts to slow at 20, accompanied by a rise in decision caution and slower non-decisional processes. The study also found that this trend continues up to 80.
	
\subsection*{Hierarchical Drift Diffusion Model}
	
Over the past few years, Bayesian analysis has gained immense popularity, especially hierarchical Bayesian analysis, which is now considered the preferred method for model-based inference in cognitive modeling \cite{lee2014bayesian},\cite{pedersen2017drift}. In hierarchical Bayesian analysis, parameter estimates are represented as probability distributions rather than fixed values. Considering prior information and the observed data, the posterior distribution represents our updated belief about the parameter values. This distribution captures the degree of uncertainty or variability in the parameter estimates, providing a more comprehensive understanding of the range of possible parameter values \cite{gelman1995bayesian},\cite{kruschke2014doing}. 
	
We tested various models to determine the best approach, including fast-dm, EZ method, and Bayesian Estimation of the Drift-Diffusion Model (HDDM). After careful exploration, we concluded that HDDM produced the most accurate results. HDDM is competent in handling slow contaminants caused by inattention or distractions, even within normal limits. Additionally, it performs well with smaller data sets. \cite{ratcliff2015individual}.
	
\section*{Method}
\subsection*{Participant}
Fifty-four participants ($31$ females; age: M= $40.4259$ years, SD= $14.7681$; rang $21-82$) were involved in this experiment. Participants were recruited from diverse groups, with 10 participants in each of the five different age ranges. The education level ranges from less than a high school degree to a graduate degree. All participants had no previous expertise, no history of mental illnesses, no severe physical disorders like Parkinson's and Huntington's, no record of visual impairment or surgery, and did not have drug abuse. The procedure was compiled with Iran University of Medical Science, underscoring its commitment to conducting research by ethical guidelines and principles,  and all participants consented before the investigation.
	
\subsection*{Materials \& Procedure}
In this study, a pre-test phase was conducted to measure the level of initial familiarity with the stimulus. During the pre-test phase, participants were exposed to the random dot motion (RDM) task to assess their baseline performance and familiarity with the stimulus.

Following the pre-test phase, an adaptive training phase was implemented. This phase aimed to enhance participants' performance(Reaction time, Accuracy) and learning in the RDM task.

After the completion of the adaptive training phase, a post-test phase was conducted to measure the subsequent growth in performance. The post-test phase assessed participants' performance (Reaction time, Accuracy) improvement after undergoing the adaptive training.

By conducting a pre-test phase to measure initial familiarity, an adaptive training phase to enhance performance, and a post-test phase to measure subsequent growth, this study aimed to investigate the effects of training on participants' performance in the RDM task and assess the extent of learning and improvement achieved through adaptive training.

In the pre-test and post-test phases, the direction and coherence of the dot motion stimuli were predetermined and consistent for all participants. This ensured that all participants experienced the same order of stimuli during these phases, allowing for standardized comparisons of performance.

However, during the adaptive phase, the trial direction and coherence levels were also predetermined, but they varied from person to person based on the individual's performance and progress.

During the RDM task and throughout the different phases of the study, all text feedback provided to participants was in the Persian language to ensure their full comprehension and understanding. This feedback, displayed in a uniform white color, consisted of text and symbols designed to convey information about participants' performance. The use of a consistent white color for the visual feedback aimed to minimize distractions and allow participants to focus on the content without any color-related biases or influences.

The visual stimulus used in this experiment was implemented with MATLAB Psychtoolbox 3.0.19. The study also included the administration of a Subjective Mental Effort Questionnaire (SMEQ) after every block of the random dot motion (RDM) task in each phase.

The SMEQ was used to assess participants' subjective perception of mental effort exerted during the task. After completing each block of the RDM task, participants were asked to rate their perceived mental effort on a questionnaire. The questionnaire likely included items that participants rated on a scale to indicate the level of mental effort they experienced during the task. This subjective measure aimed to capture participants' perceived cognitive load and mental engagement throughout the experiment.

By including the SMEQ after each block of the RDM task in every phase, the study sought to gain insights into participants' perceived mental effort and examine potential variations in cognitive load throughout the different phases of the experiment. This information could provide valuable insights into participants' subjective experiences and their engagement with the task at different steps of the study.

In the design of test phases and questionnaires, the results of the previous versions of the stimulus as well as the corrections made in them have been used. By analyzing the results of the previous versions of the stimulus and identifying their weaknesses and defects, efforts have been made to apply the necessary improvements in the final version of the test phases and questionnaires.

\subsection*{Pre-test and Post-test Phase}
	The pre/post-test phase aimed to assess the initial familiarity level and the growth level after phase training. It consisted of a single block comprising $120$ trials. The coherence number was approximately equal, generated from a uniform distribution, which may not be suitable for personal comparisons but was deemed acceptable for the study. There were no specific criteria for passing this phase.
	
	No specific conditions were imposed for participants to pass this phase. The standard parameters for the dots aperture, as described in Kiani et al. (2013) \cite{kiani2013integration} were employed. During the pre/post-test phase, neutral auditory feedback was provided to inform the participants that their responses had been recorded.
	
	The inclusion of the pre/post-test phase was important for understanding the baseline performance of participants before any training intervention. It provided a reference point for assessing the effectiveness of the subsequent training phase and determining the extent of growth or improvement in visual perception and decision-making abilities.
	
	By measuring performance both before and after the training phase, it was possible to evaluate the impact of the training task on participants within each age range. The pre-test phase established a baseline against which the effects of training could be assessed, while the post-test phase allowed for the detection of any changes or improvements resulting from the training intervention. This design enabled a direct comparison of each participant's performance across the two phases.
	
	  	\begin{figure}[H]
		\centering
		\includegraphics[width=0.8\linewidth]{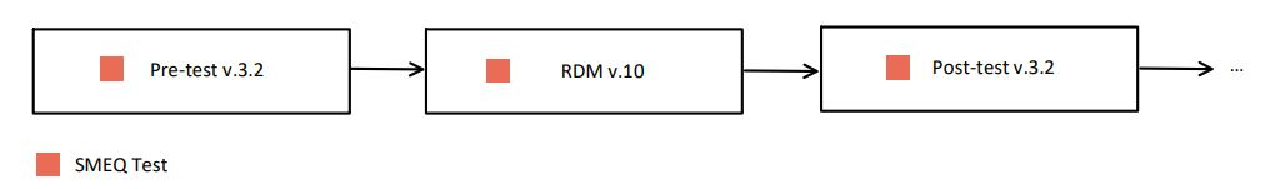}
		\caption{Pre- and post- test procedure.}
		\label{fig: 9}
	\end{figure}
	
\subsection*{Adaptive Training Phase}
	The primary objective of this phase was to facilitate and expedite the training process for participants across different age groups. To achieve this, the stimulus was carefully designed to start from an easy and comprehensible state and gradually adapt to the standard stimulus over 10 steps. Each step involved specific restrictions on participant accuracy and reaction time.
	
	The dot parameters, including size, coherence, and aperture, were gradually adapted to the standard stimulus over the 10 steps.(see table)The adaptation process ensured that the stimulus became progressively more challenging and closely aligned with the standard stimulus. The density of the dots remained constant at $16.7$, which matched the standard stimulus.\cite{kiani2013integration}
	
	Based on previous versions of the study, it was determined that the speed of adaptation did not significantly impact the results. Therefore, the speed of adaptation was kept constant, matching that of the standard stimulus.
	
	The maximum number of trials per block was set at $220$, and it was possible to conduct up to three iterations (blocks) of the training phase. The minimum number of trials required was $130$, while the maximum number of trials reached $660$.
	
	For each step, the coherence and direction of the dots were predetermined. During the predetermined trials, every $5$ consecutive trials ($6$ for stages $9$ and $10$), all coherences of that stage were included. However, depending on the participant's performance, the order of presentation did not necessarily follow this sequence. The minimum number of trials for each step ensured that participants encountered a rich combination of coherences and directions. Regardless of their performance, every participant was required to respond to this minimum accepted number of trials for each step.
	
	Additionally, this phase included limitations on the mean reaction time and accuracy for each step. Participants were expected to achieve an accuracy above $80\%$ and a mean reaction time below the maximum accepted value for each step. These additional conditions were assessed after the minimum trials were completed. If the participant did not meet these conditions, trials continued until both accuracy and reaction time requirements were satisfied. If a participant could not pass all steps within $220$ trials, they took a short rest and started the next block, beginning from the first step. This cycle continued until the participant successfully passed all steps or reached $660$ trials, which marked the end of block $3$. Afterward, the participant was ready to proceed to the next phase following a brief $~5$ minute rest or could choose to abort from the experiment. If a participant reached the last stage but narrowly missed satisfying the additional conditions, it was still considered acceptable.
	
	Throughout this phase, various visual and auditory feedback mechanisms were employed based on the efficiency observed in previous versions of the study. (see figure )
	
	visual feedback was provided to participants to enhance their learning and performance. At the end of each successfully passed step, text feedback was displayed on the screen, acknowledging their achievement. To ensure timing compliance, participants received Mean RT (Reaction Time) feedback. This feedback encourage them to reduce their reaction time and It was repeated after five consecutive trials until the related condition was not satisfied. appeared for $3$ seconds at the center of the display.
	
	After each trial, correctness feedback was presented in the form of symbolic check and cross feedback. This visual feedback appeared for $0.5$ seconds at the center of the display. The check symbol indicated a correct response, while the cross symbol represented an incorrect response.
	
	In addition to visual feedback, auditory feedback which is a distinctive beep sound was played to indicate whether the participant's response was correct or incorrect.
	
	Along with the auditory feedback, symbolic feedback was displayed on the screen after each trial. This symbolic feedback  This feedback served as an additional cue to reinforce the participant's understanding of their performance.
	
	The combination of visual and auditory feedback in this phase aimed to provide comprehensive and timely information to participants, helping them monitor their progress, make necessary adjustments, and reinforce their learning and reduce their reaction time throughout the training process.
	
    \begin{figure}[H]
		\centering
		\includegraphics[width=\linewidth]{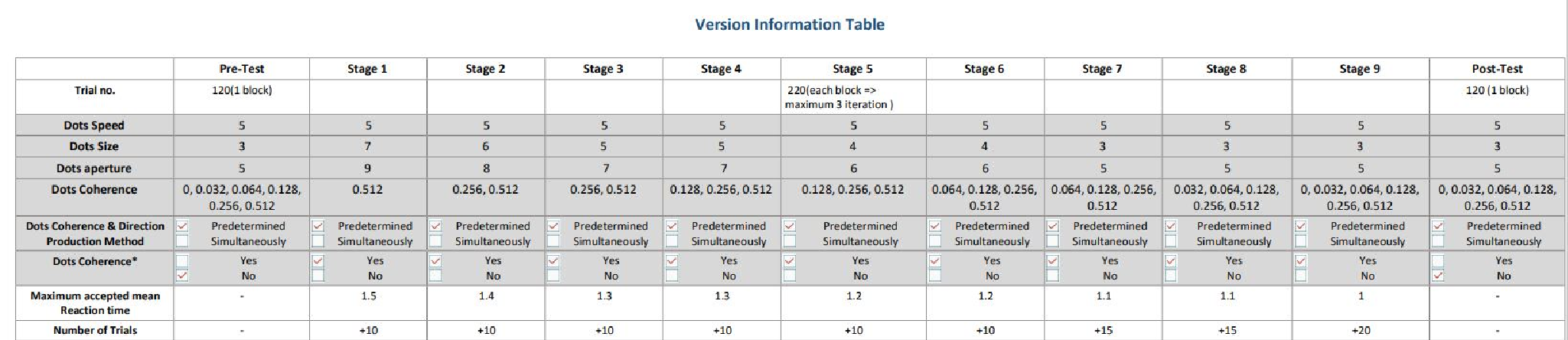}
				\caption{Adaptive training phase procedure.}
		\label{fig: 10}
	\end{figure}

	\subsection*{Data Preprocessing}
	
\begin{figure}[H]
		\centering
		\includegraphics[width=0.9\linewidth]{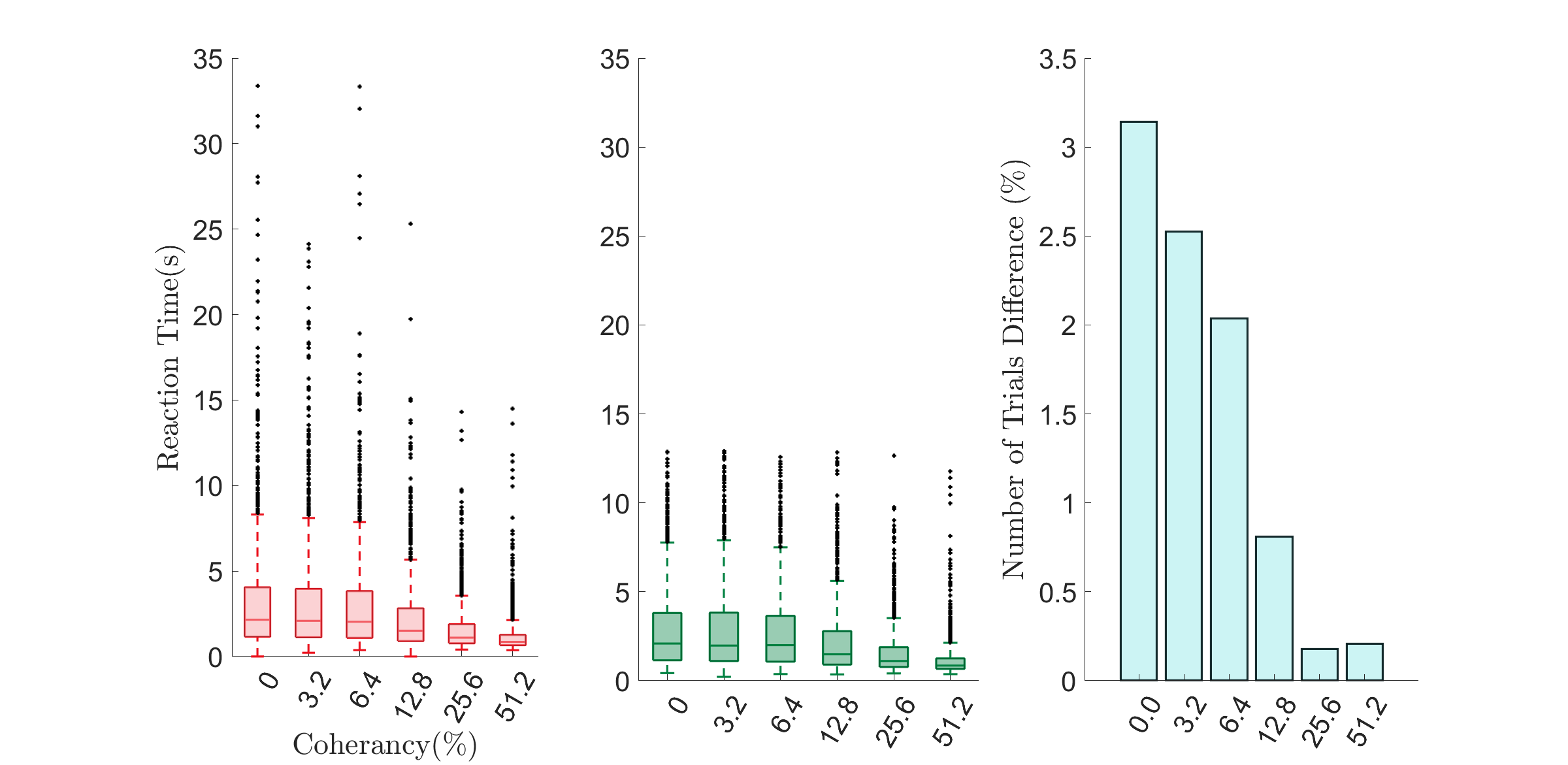}
		\includegraphics[width=0.9\linewidth]{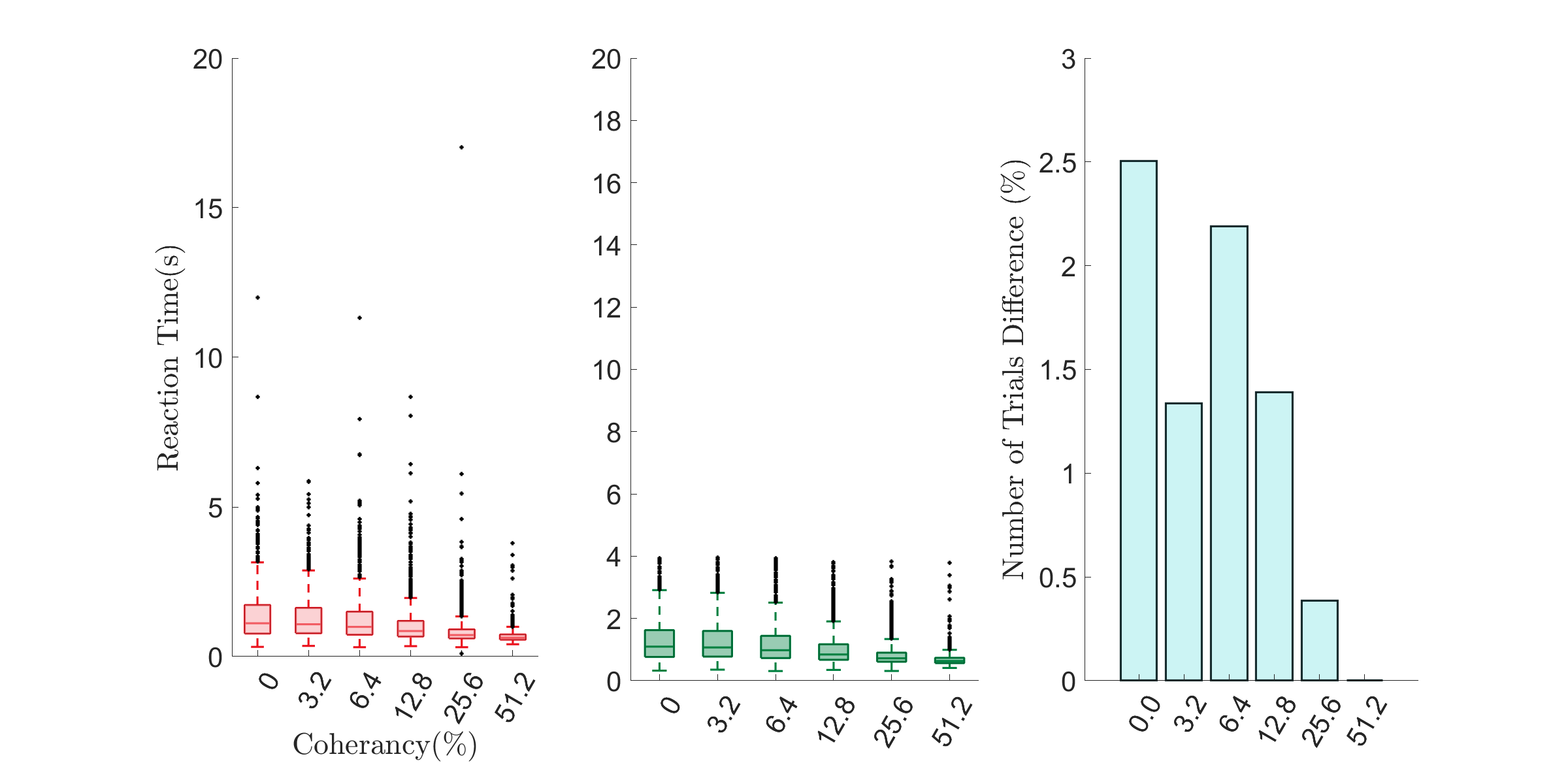}
		\caption{Data pre-processing. Top: pre-training, Bottom: post-training. In each image, right to left: raw data with outliers, without outliers, and the percentage of trials eliminated in each coherency.}
		\label{fig: S.1}
	\end{figure}

	First, we eliminate trials with RTs $3.5$ standard deviation (SDs) above the average RTs in each condition \cite{dix2020incentive}. Also, trials with TRs below $0.2s$ are removed as they are much faster than usual responses. By this criteria, $1.6\%$ of trials in pre-training and $1.33\%$ in post-training were removed. To assess whether the number of trials extracted from each coherency level is evenly distributed, we compared the percentage of outlier trials in each coherency level for both the pre-and post-training data (figure \ref{fig: S.1}). It is worth noting that the number of outliers is significantly higher for lower coherence levels due to the varying levels of task difficulty across subjects. Additionally, post-training data exhibits fewer outliers, particularly for lower coherence levels, indicating that subjects could respond to the task with reasonable reaction times. 
	
\subsection*{Hierarchical Drift diffusion model}
	The diffusion model fits each subject using HDDM, Python toolbox, version $0.6$ \cite{wiecki2013hddm}, which works based on hierarchical Bayesian estimation. Unlike traditional diffusion models, the Hierarchical Bayesian model considers inter-individual variability. Each subject parameter is estimated from a whole group distribution or samples randomly selected from larger populations. This approach is vital for most cognitive tasks in which providing enough sample data per subject (or per condition) is nearly impossible. Also, some model’s parameters can remain constant for all individuals (or in a condition), and others vary between subjects \cite{vandekerckhove2011hierarchical}. Another advantage is that the HDDM model estimates the posterior distribution of parameters rather than only a value,  which implies the parameters’ uncertainties. In our work, we consider seven variants of HDDM models, depending on whether three main parameters, boundary (a), drift rate (v), and non-decision time (tER), depend on the task conditions and trials’ coherency. The bias (z) and inter-trial variability remain constant and are not considered during the estimation process. Each subject’s model parameters are estimated using $2000$ posterior samples, $100$ burn-in, and five percent outliers. In addition, the posterior distribution of parameters is used to compare their alternation after training. We assess models regarding the deviance information criterion (DIC) to evaluate how well the posterior model might anticipate future observations \cite{zhang2014dissociable}. As less DIC is, the model estimation is better. Finally, we used t test to statistically determine whether the training had affected the model’s parameters.
	
	In our best model (the one with the lowest DIC value), all parameters (a, v, and tER) are based on the condition (\ref{table 1}, model 7). To further analyses, we call this our first model. We also consider another model as well as the first, despite the drift rate depending on the coherency and conditions. The model implies how training changes the participants’ performance at different task levels. The task with higher coherency is considered hard, and an easy level has low random dots coherence. This model is called the second model in our later analyses. In addition, we consider participants’ ages, divide them into two groups: young (younger than average) and old.
	
	The fitting quality is evaluated by comparing the histogram and quantile of the model-regenerated data with our empirical data. The histograms of reaction time (RTs) in correct and false trials are shown separately, and $0.1, 0.5,$ and $0.9$ quantiles are considered for estimated and empirical data.

	\setlength{\tabcolsep}{10pt} 
	\renewcommand{\arraystretch}{1.5}
	\begin{table}[h!]
		\begin{adjustbox}{width=\columnwidth,center}
	     	\centering
		   \begin{tabular}{|c|c|c|c|c|c|c|c|}
			\cline{2-8}
			\multicolumn{1}{c}{ } & \multicolumn{2}{|c}{a} & \multicolumn{2}{|c}{v} & \multicolumn{2}{|c|}{t} & \multirow{2}{*}{DIC}  \\ 
			\cline{2-7} 
			\multicolumn{1}{c|}{} & Condition& Coherency& Condition& Coherency& Condition& Coherency & \\
			 \cline{1-8}
			 Model 1 & & &+& & & &37731.8408 \\
			 Model 2&  & &  & & +& &36725.9529 \\   
			  Model 3& & & + & & + & &36367.5913 \\ 
			  Model 4& + & & & & & &34703.7134 \\   
			  Model 5& + & & + & & & &34491.4561 \\ 
			  Model 6& + & & & & +& &34378.2644 \\ 
			  Model 7& + & & + & & +& &34161.2076 \\ 
			 \textcolor{red}{Model 8}& + & & +& +&+ & & 28051.9062 \\               
			\bottomrule
			
	     	\end{tabular}
	   \end{adjustbox}
    		\caption{HDDM Parameters’ Statistics, pre and post training}
             \label{table 2}
	\end{table}
	
\section{Result}
\subsection*{Learning Effects on Behavioral Performance}
\begin{figure}[H]
		\centering
		\includegraphics[width=\linewidth]{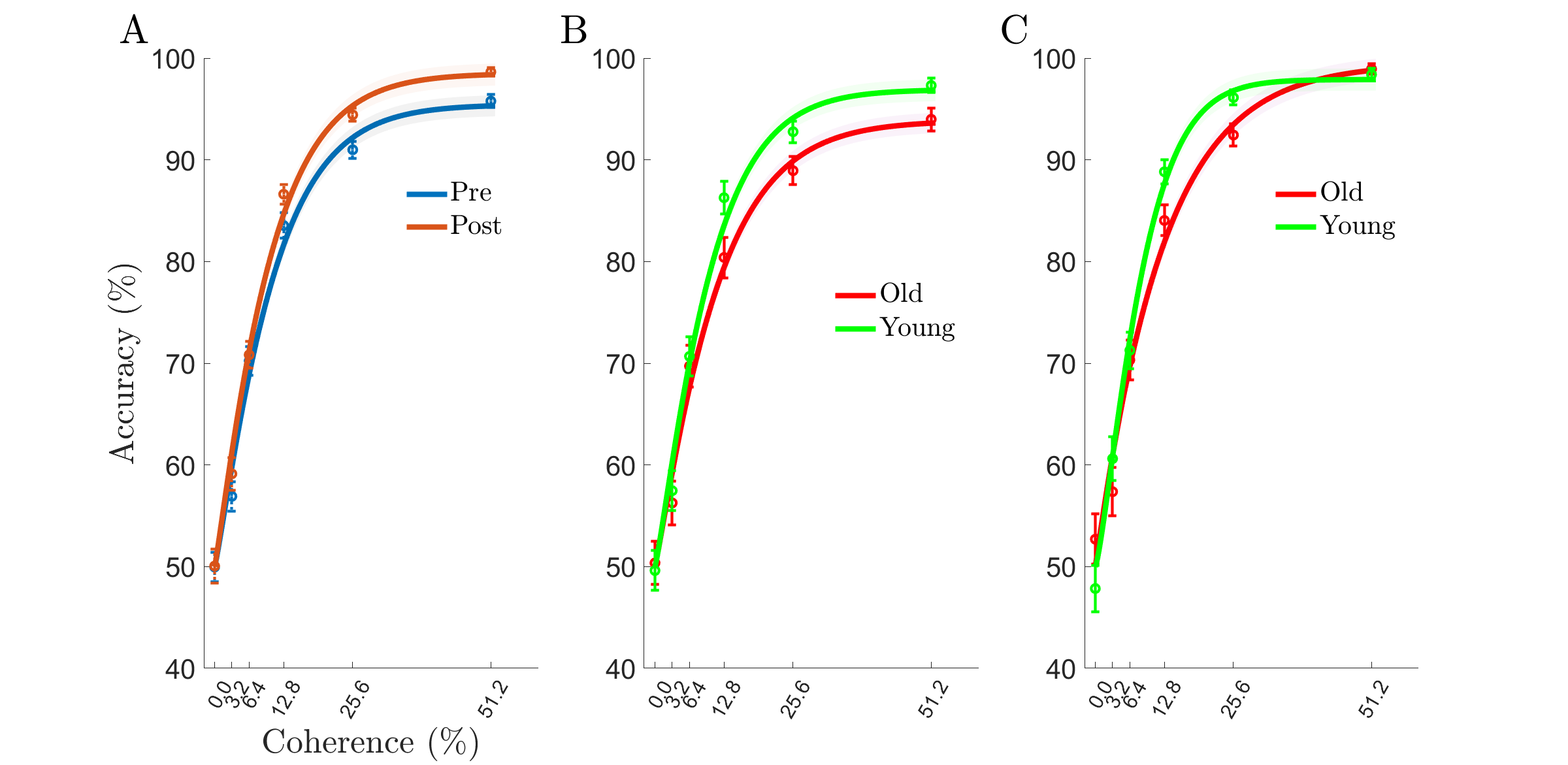}
		\caption{Behavioural Results: Psychometric Curve, whole subjects for pre- and post-training (A), pre-training for old and young(B) and after training for young and old group (C).}
		\label{fig: 1}
	\end{figure}

\begin{figure}[H]
	\centering
	\includegraphics[width=\linewidth]{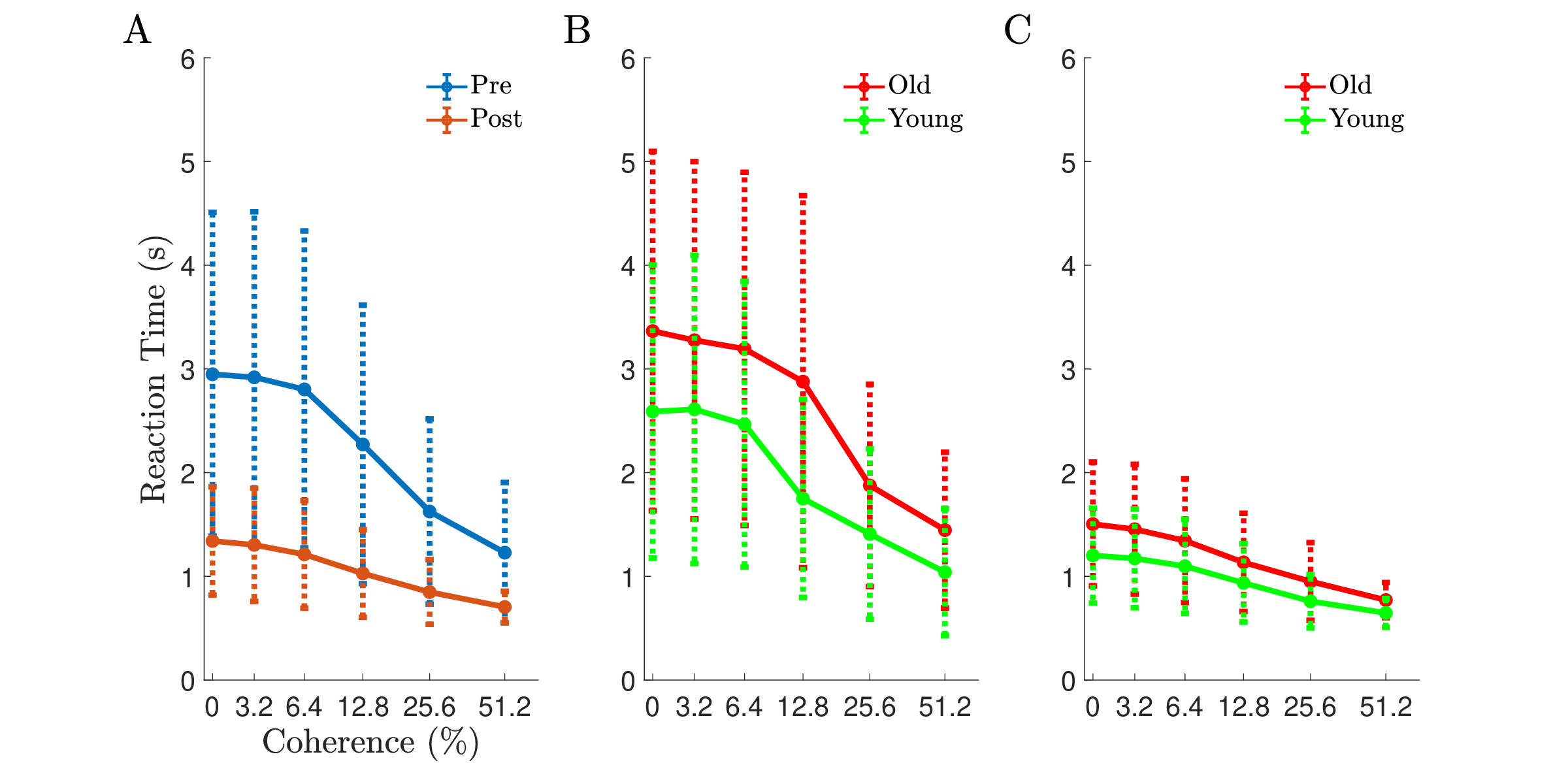}
	\caption{Behavioural Results: Chronometric Curve, whole subjects for pre- and post-training (A), pre-training for old and young(B) and after training for young and old group (C).}
	\label{fig: 2}
   \end{figure}

	Accuracy is defined as a number of correct-answered trails over the number of all trails in a task. To compare participants’ accuracy and reaction time, psychometric and chronometric functions are used (figure \ref{fig: 1},\ref{fig: 2}). These curves represent the relationship between a particular stimulus feature (e.g. coherency) and subjects’ task responses (e.g., accuracy and reaction time). It is clear that the training significantly affected participants' performance.  Generally, subjects’ accuracy increased in post-training (pre: mean: $75.5081\%$, range: $[73.0124\%, 80.7531\%]$, Post: mean: $79.0361\%$, range: $[77.2524\%, 84.7162\%]$) and their RTs reduced significantly (pre: mean: $2.3248s$, range: $[0.5801s,5.1828s]$, Post: mean: $1.0670s$, range: $[0.5686s, 1.7578s]$).
	
	Regarding age analyses, we compare the young and the old psychometric and chronometric curves are compared in each condition (figure \ref{fig: 1},\ref{fig: 2}). Both before and after training the main effects of age were negligible (Accuracy: pre: young: $43.1764\%$, old: $34.5784\%$, post: young: $45.1484
	\%$, old: $36.2210\%$. RTs: pre: young: $1.3683s$, old: $1.0326s$, post: young: $
	0.6177s$, old: $0.4836s$.).
	
\subsection*{Hierarchical Drift Diffusion Model Results}
\begin{figure}[H]
    	\centering
    	\includegraphics[width=\linewidth]{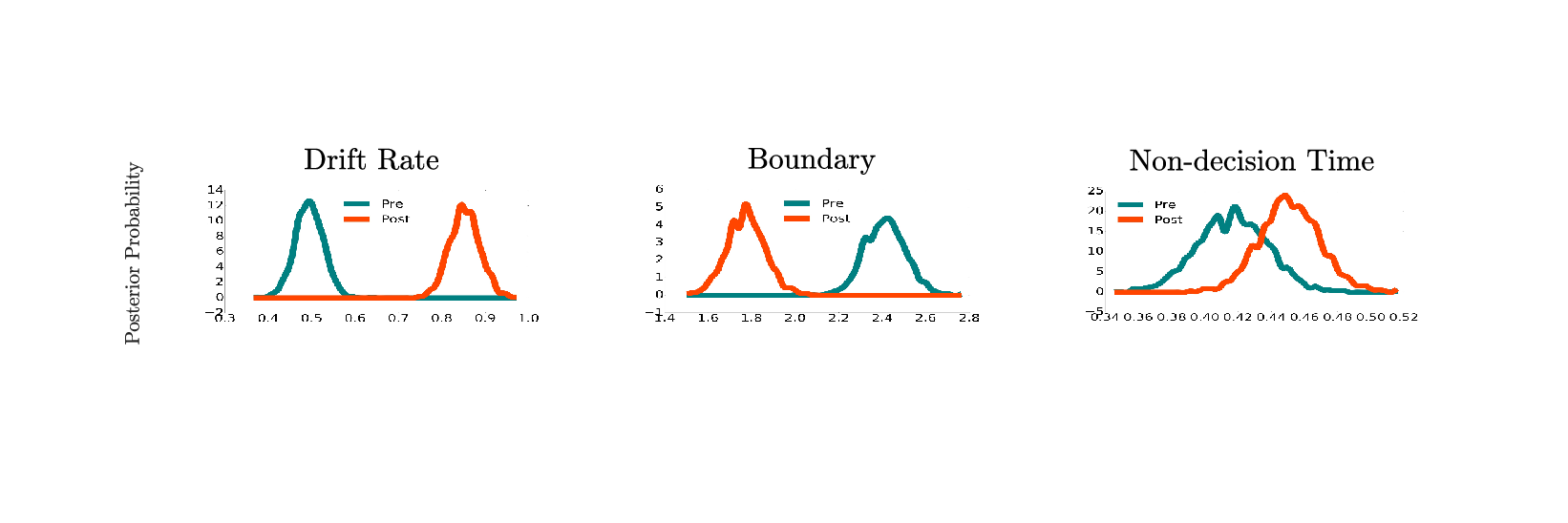}
    	\caption{hDDM Parameters: right to left: Drift rate, decision boundary and non-decision time in pre training (red) and post training (green).}
    	\label{fig: 3}
    \end{figure}

	In the first model, all parameters only depend on the condition. To evaluate which parameters lead to accuracy improvements and reaction time reduction, we compare their posterior distribution between pre and post-training (figure \ref{fig: 3}). While the drift rate increases noticeably after training, the bound of decision dramatically decreases. It means that the evidence accumulation rate rises, and the participants assume a lower decision-making threshold for their responses. Moreover, non-decision time differences are not significant in each condition, so the decrease in reaction time in post-training is not due to time in perceptual encoding and motor movement. The summary of the parameters' statistics is shown in tabel \ref{table 2}.

\begin{figure}[H]
  	\centering
  	\includegraphics[width=0.8\linewidth]{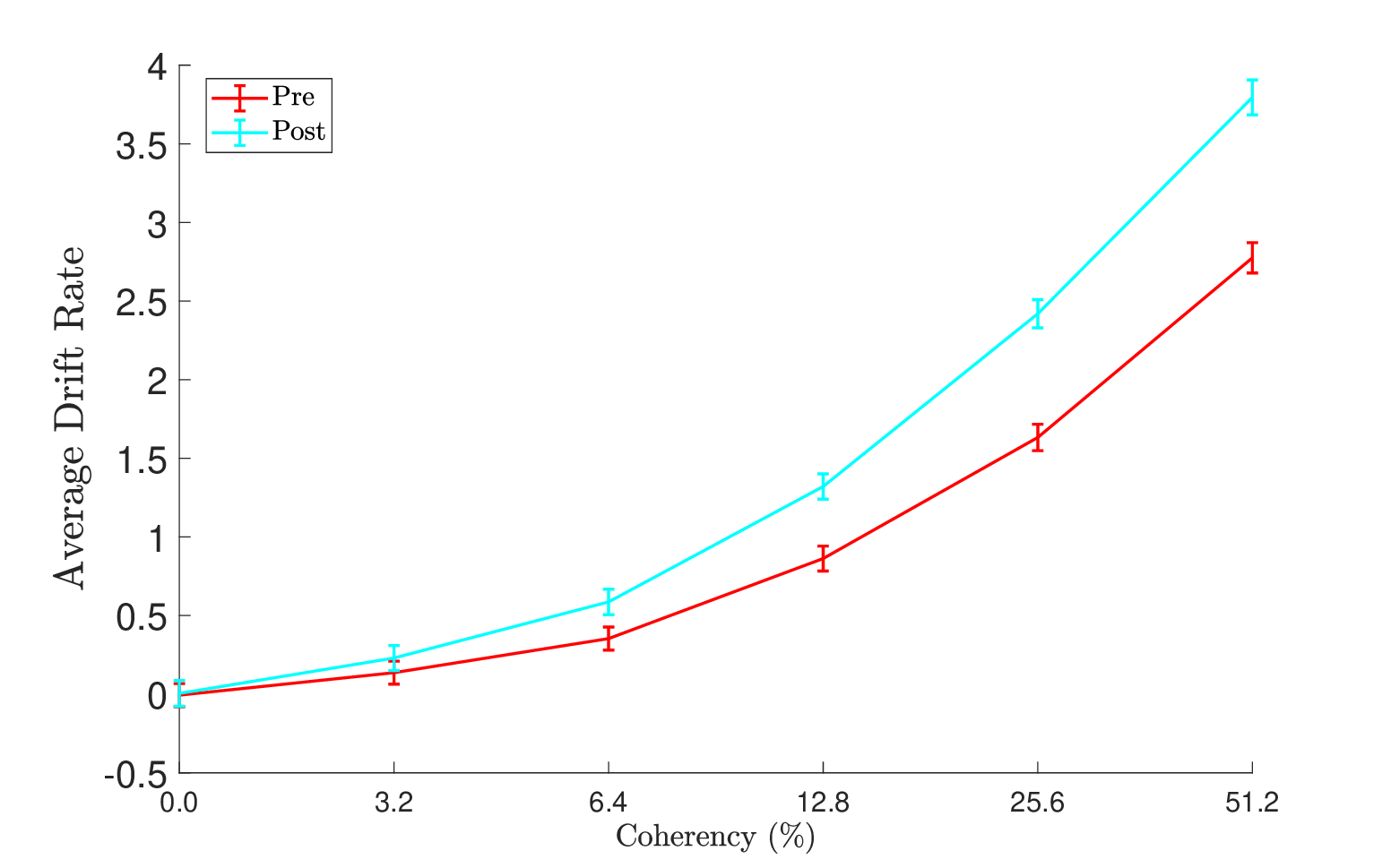}
  	\caption{Average drift rate among participants in each coherency level for pre- and post-training.}
  	\label{fig: 8}
  \end{figure}

	The second model is the same as the first model, in spite of drift rate which depends on either coherency and condition. The model shows that training increases drift rate in all trial levels (high or low coherence trials), and this increment is higher for easier trials (figure \ref{fig: 8}). As we expected, the evidence accumulation is slower for hard and ambiguous tasks, like trials with lower coherency. 
	
\begin{figure}[H]
   	\centering
   	\includegraphics[width=\linewidth]{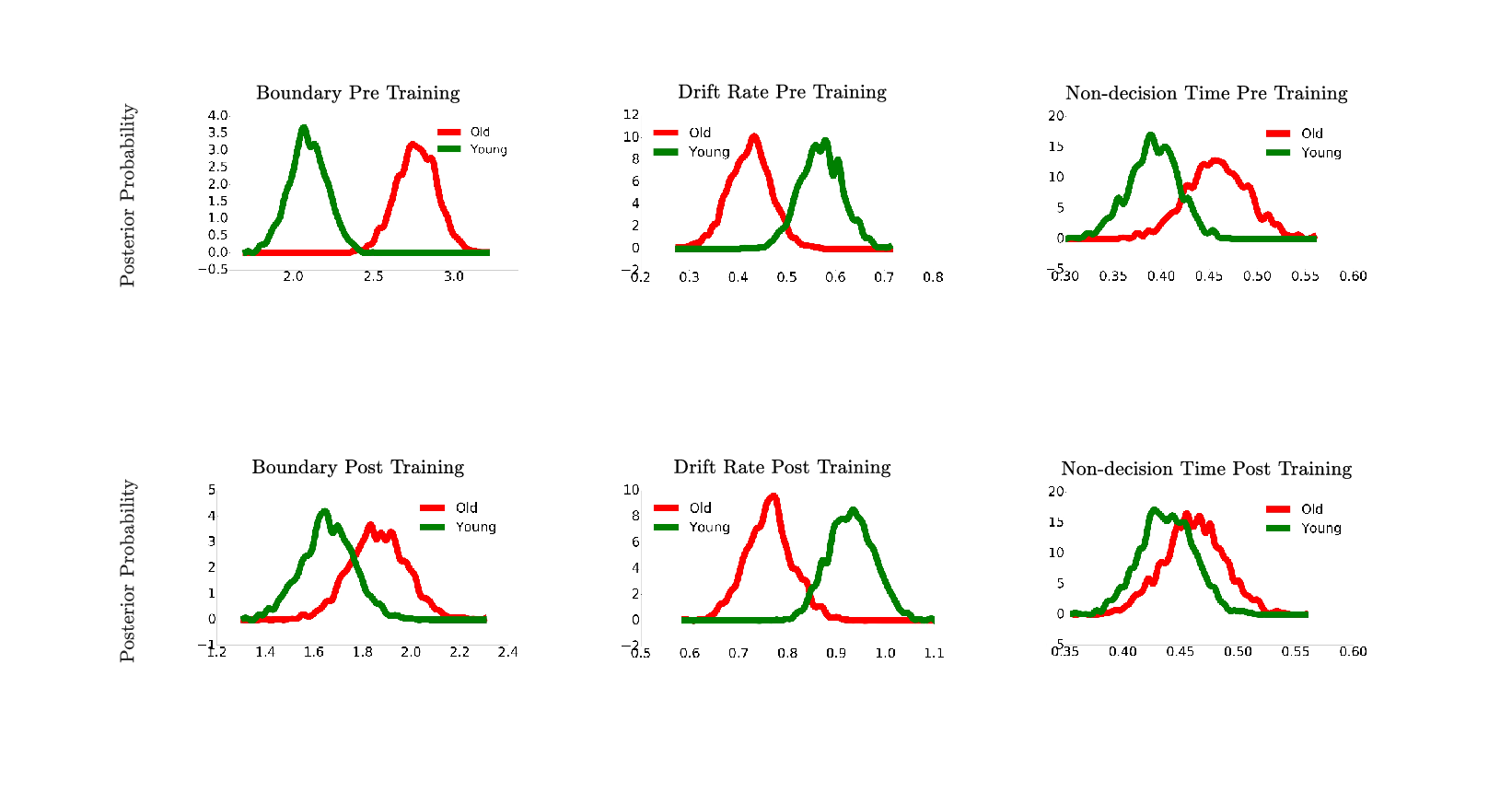}
   	\caption{Model parameters among old and young group, in two conditions.}
   	\label{fig: 4}
   \end{figure}

\begin{figure}[H]
   	\centering
   	\includegraphics[width=\linewidth]{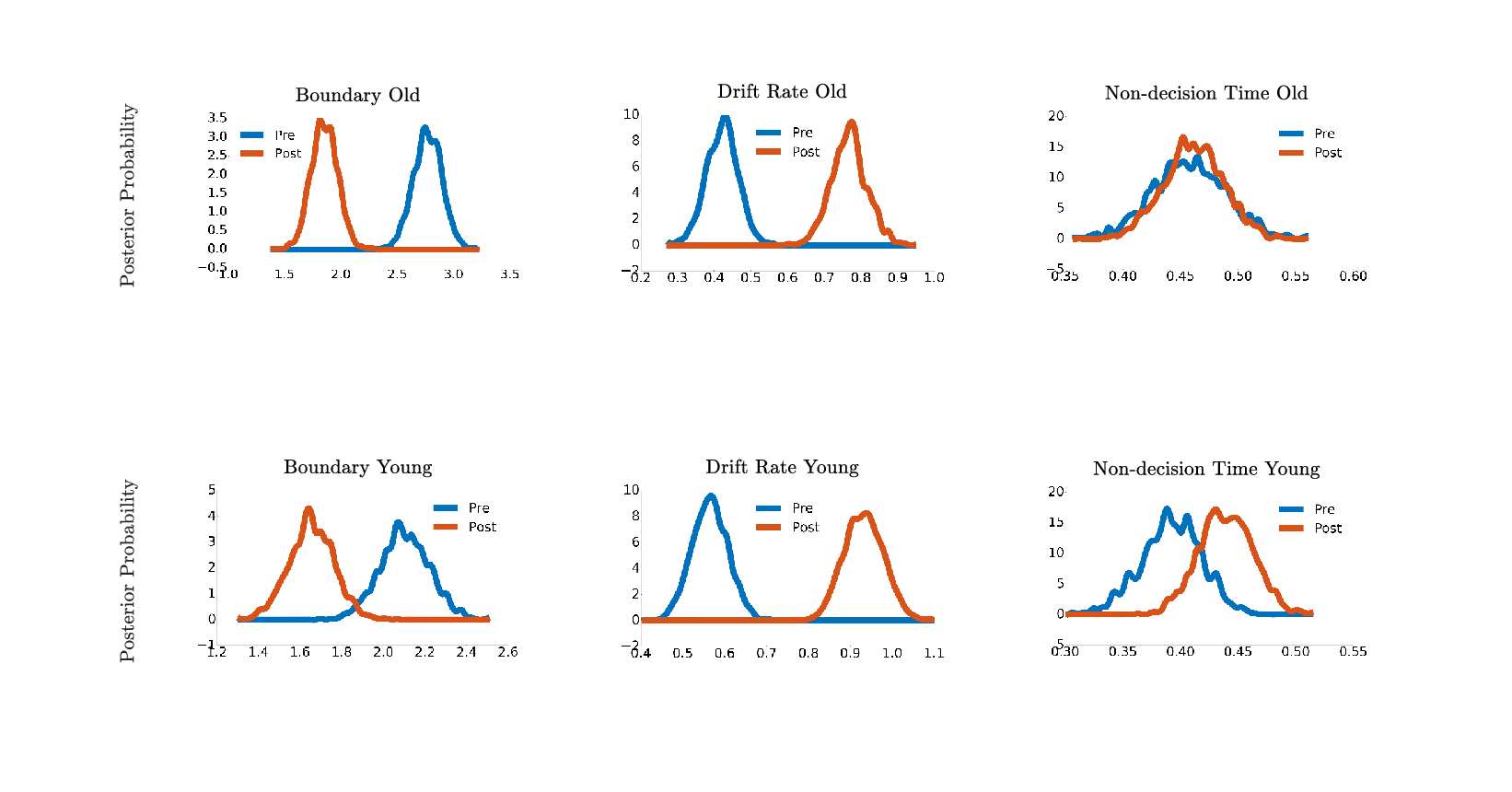}
   	\caption{Model parameters in pre and post condition, for two groups.}
   	\label{fig: 5}
   \end{figure}

	In addition, to analyses the age effects on learning procedure and decision processing, posterior distribution of the first model’s parameters are compared between groups (figure \ref{fig: 4}) and within groups before and after training (figure \ref{fig: 5}). In pre-training, the old respond to the task with a higher boundary in comparison to the young group, but the difference is not noticeable after training. Moreover, the young have a higher drift rate, and the non-decision time is nearly the same for both groups in both conditions. 
	
	\setlength{\tabcolsep}{10pt} 
	\renewcommand{\arraystretch}{1.5}
	\begin{table}[h!]
		\centering
		\begin{tabular}{|c||c|c|c|c|}
			\hline
			Parameters & Condition & Mean & Std & P-value\\ [0.5ex]
			\hline\hline
			\multirow{2}{*}{Boundary} & Pre & 2.40295 & 0.0949367 & \multirow{2}{*}{1.3e-10} \\
			                          & Post & 1.77968 & 0.0864387 &  \\\hline
			                          
		    \multirow{2}{*}{Drift Rate} & Pre & 0.498486 & 0.0317933 & \multirow{2}{*}{1.75e-17} \\
			                           & Post & 0.84889 & 0.0317933 &  \\\hline
			
			\multirow{2}{*}{Non-decision Time} & Pre & 0.418694 & 0.0198375 & \multirow{2}{*}{0.68} \\
			                                   & Post & 0.451716 & 0.0184613 &  \\                        
			\hline

		\end{tabular}
		\caption{HDDM Parameters’ Statistics, pre and post training}
		\label{table 1}
	\end{table}
	
	\subsection*{Quality of Fits}
\begin{figure}[H]
   	\centering
   	\includegraphics[width=\linewidth]{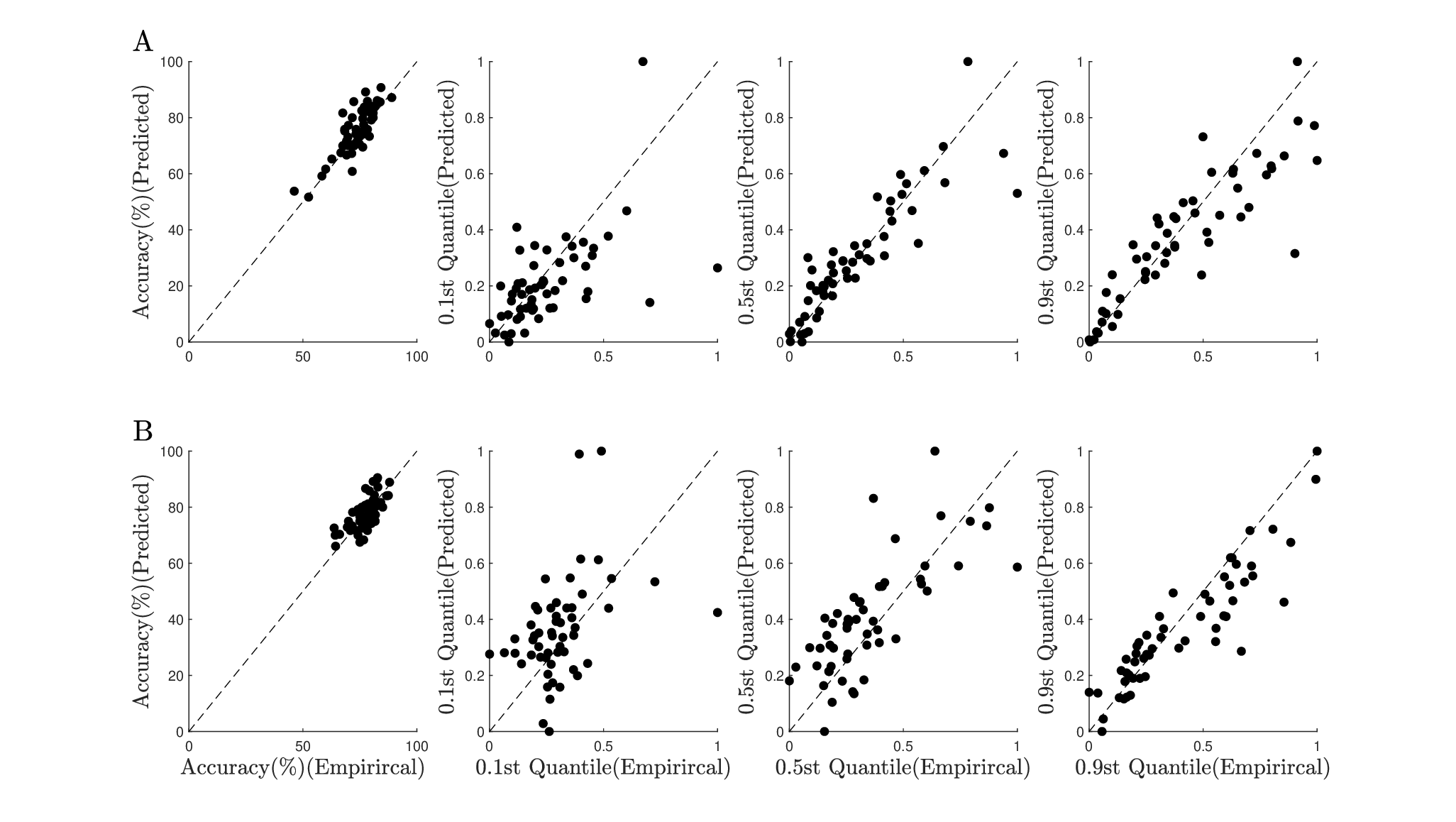}
   	\caption{A: Accuracy and $0.1$, $0.5$, and $0.9$ RT quantile for pre-training data and predicted value (model data). Right to left: Accuracy (corr: $82.39\%$, P-value:$1.9855e-14$), $0.1$ quantile (corr: $55.5394\%$, P-value:$1.3059e-05$),  $0.5$ quantile (corr: $88.8183\%$, P-value:$3.3852e-19$),  $0.9$ quantile (corr: $87.9229\%$, P-value:$2.2385e-18$). B: Accuracy and $0.1$, $0.5$, and $0.9$ RT quantile for post-training data and predicted value (model data). Right to left: Accuracy (corr: $69.599\%$, P-value: $5.1422e-09$), $0.1$ quantile (corr: $38.8612\%$, P-value: $0.0036$),  $0.5$ quantile (corr: $75.5991\%$, P-value: $3.8691e-11$),  $0.9$ quantile (corr: $90.5932\%$, P-value: $4.7343e-21$)}
   	\label{fig: 6}
   \end{figure}

	The model parameters are valid and interpretable if only the model regenerated data fits the behavioural data. The empirical value of accuracy and three quantiles are plotted against model data in figure \ref{fig: 6}. Most values seem aligned with the perfect congruence line, and the model could predict behavioural processes. 
	
\section*{Discussion}
This study explores how learning affects the decision-making process between young and old age groups. A latent sensory processing model involving a diffusion model with three key parameters - boundary, drift rate, and non-decision time - has been used to understand the amount of evidence required to make a decision, the speed of evidence accumulation during the decision-making process, and the time needed for stimulus processing and motor movement. In general, participants responded more effectively to the task with a lower decision bound and a higher drift rate. However, there was very little difference in non-decision time before and after the training sessions. Moreover, older participants took longer to accumulate sensory evidence and had a higher decision boundary before the training sessions, but they were able to reduce their decision boundary and increase their drift rate after the training sessions.

	\newpage
	\section*{Images}
	
   \begin{figure}[H]
   	\centering
   	\includegraphics[width=\linewidth]{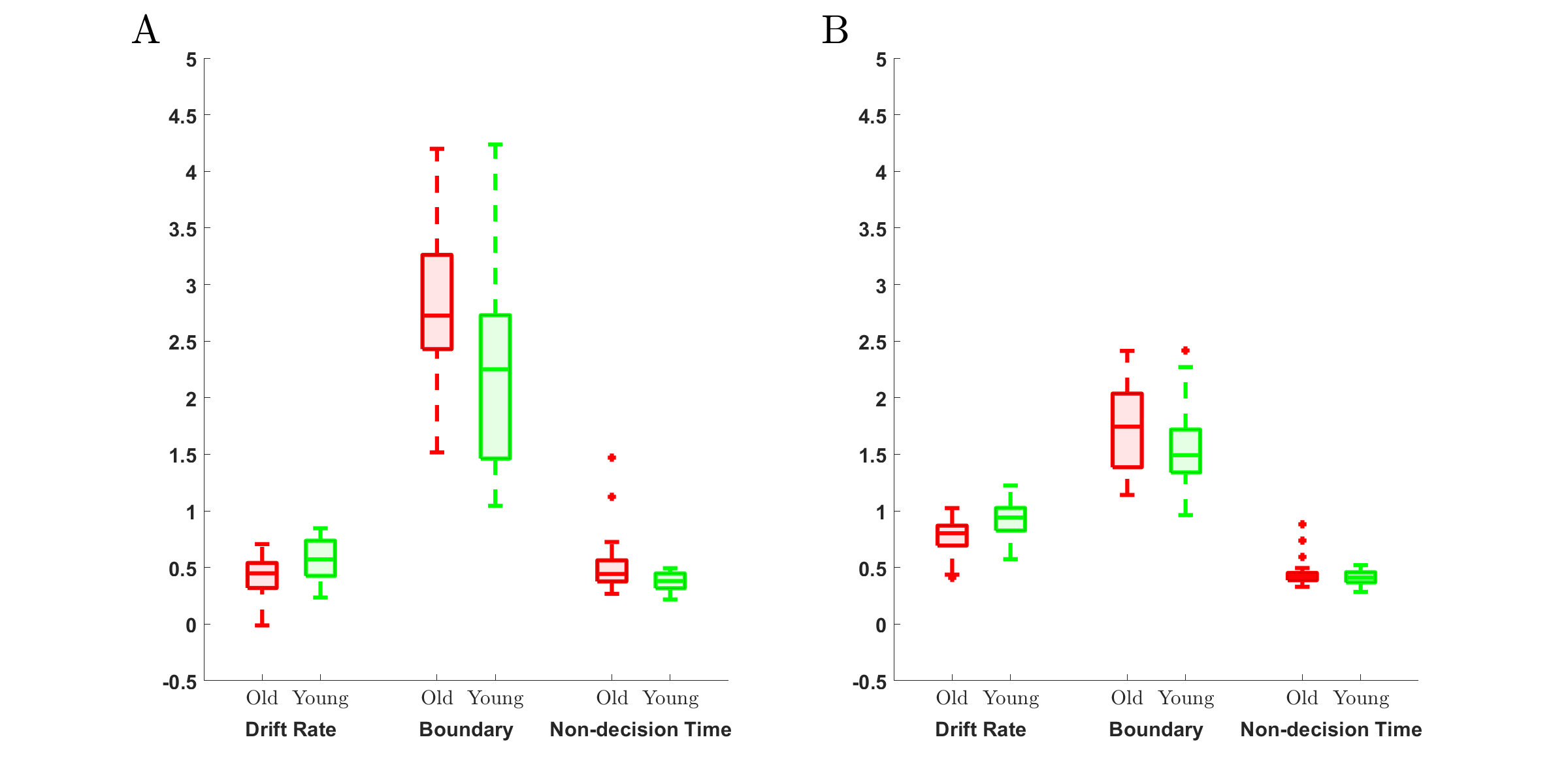}
   	\caption{HDDM parameters between the old and young. Right: pre-training (drift rate: $0.00433$, boundary: $0.0212$, non-decision time: $0.01$), Left: post-training (drift rate: $0.000433$, boundary: $0.0189$, non-decision time: $0.15$)}
   	\label{fig: 7}
   \end{figure}

	\newpage

	\section*{Supplementary}
	
	\newpage
	\bibliographystyle{plain}
	\bibliography{references}

\end{document}